%% file: supermanifest_v5.tex
\documentclass[11pt,a4paper]{article}
\usepackage[latin1]{inputenc}
\usepackage[totalwidth=410pt, totalheight=590pt]{geometry}
\usepackage{cite}

\usepackage{amsmath}
\usepackage{amsfonts}
\usepackage{amssymb}

\numberwithin{equation}{section}

\input{macros}

\begin{document}

 \pagestyle{empty}
% \vskip-10pt
% \hfill {\tt hep-th/0602193}

\begin{center}

\vspace*{2cm}

\noindent
{\LARGE\textsf{\textbf{Manifest superconformal covariance \\[5mm] in six-dimensional $(2,0)$ theory}}}
\vskip 3truecm

{\large \textsf{\textbf{P\"ar Arvidsson}}} \\
\vskip 1truecm
{\it Department of Fundamental Physics\\ Chalmers University of
  Technology \\ SE-412 96 G\"{o}teborg,
  Sweden}\\[3mm] {\tt par.arvidsson@chalmers.se} \\
\end{center}
\vskip 1cm
\noindent{\bf Abstract:}
A superconformal generalization of Dirac's formalism for manifest conformal covariance is presented and applied to the free $(2,0)$ tensor multiplet field theory in six dimensions. A graded symmetric superfield, defined on a supercone in a higher-dimensional superspace is introduced. This superfield transforms linearly under the transformations of the supergroup $OSp(8^*|4)$, which is the superconformal group of the six-dimensional $(2,0)$ theory. We find the relationship between the new superfield and the conventional $(2,0)$ superfields in six dimensions and show that the implied superconformal transformation laws are correct. Finally, we present a manifestly conformally covariant constraint on the supercone, which reduces to the ordinary differential constraint for the superfields in the six-dimensional space-time.

\newpage
\pagestyle{plain}

%\tableofcontents

\section{Introduction}

In 1936, P.~Dirac~\cite{Dirac:1936} introduced a novel way of treating conformal symmetry. By considering fields living on a projective hypercone, defined in a \emph{conformal space} with two additional dimensions compared to the usual Minkowski space-time, he found a way of making this symmetry \emph{manifest}. Since then, his work has been refined and extended~\cite{Kastrup:1966,Marnelius:1979,Marnelius:1979b,Marnelius:1980}, most notably by G.~Mack and A.~Salam~\cite{Mack:1969}. The latter paper provides a careful recipe for how to reduce manifestly covariant fields on the projective hypercone to conventional space-time fields.

The purpose of the present paper is to generalize this recipe, and Dirac's hypercone formalism, to incorporate manifest \emph{superconformal} symmetry. It is well known that consistent superconformal field theories cannot exist in space-times with more than six dimensions. Moreover, the largest possible supersymmetry consistent with superconformal invariance in six dimensions is the chiral $\mathcal{N}=(2,0)$~\cite{Nahm:1978}. As a concrete example of these concepts, we have chosen to work in the context of this six-dimensional $(2,0)$ theory~\cite{Witten:1995}, or more concretely, with the free $(2,0)$ tensor multiplet field theory.

We will rely heavily on a previous publication~\cite{Arvidsson:2005} treating the superconformal symmetry of the $(2,0)$ tensor multiplet, and we will refer frequently to results obtained there. This paper also provides a more detailed description of the origins and the degrees of freedom of the theory.

The outline and the main results of the present paper are as follows: \Secref{superconformal} introduces the superconformal theory, characterized by the supergroup $OSp(8^*|4)$. This supergroup is the isometry group for a \emph{superconformal space} with eight bosonic and four fermionic dimensions, and the group transformations act linearly on the coordinates in this space. We introduce an invariant projective supercone (which is a superspace generalization of Dirac's hypercone) in the superconformal space and show how to relate this to the conventional coordinates in $(2,0)$ superspace.

In \Secref{fields}, a new graded symmetric superfield is defined, living on the supercone. We demand this field to be supertraceless and require its contraction with the superspace coordinate to vanish everywhere on the supercone; this yields an algebraic relationship to the superfields of the $(2,0)$ tensor multiplet. This relation is consistent with an interesting generalization of Mack and Salam's recipe and we find that a linear $OSp(8^*|4)$ transformation of the new superfield implies the known superconformal transformation laws for the $(2,0)$ superfields. The foundation for the new recipe is to require both translations and supersymmetry transformations to act only differentially on fields, and it should be applicable to other superconformal theories as well.

Finally, in \Secref{diff_con} we show that the differential constraint, which is necessary for the consistency of the superfield formulation in the six-dimensional space-time, may be formulated in a simple and manifestly conformally covariant way in the superconformal space.

\section{The superconformal group and the supercone}
\seclab{superconformal}

Consider a superspace with eight bosonic and four fermionic dimensions, having the natural isometry group $OSp(8^*|4)$. We will call this the \emph{superconformal space}, in analogy with Dirac's notion~\cite{Dirac:1936} of a conformal space in the bosonic case.

The fundamental (anti)commutation relations defining $OSp(8^*|4)$ are~\cite{Kac:1977,Claus:1998}
\beq
\eqnlab{superalgebra}
\Big[J_{\sss AB}, J_{\sss CD} \Big\} = -\frac{1}{2} \Big( I_{\sss BC}
J_{\sss AD} - (-1)^{\sss AB} I_{\sss AC} J_{\sss BD} -
(-1)^{\sss CD} I_{\sss BD} J_{\sss AC} +
(-1)^{\sss AB+CD} I_{\sss AD} J_{\sss BC} \Big),
\eeq
where obviously $J_{\sss AB}$ is the generator and the bracket in the left hand side is an anticommutator if both entries in it are fermionic, otherwise it is a commutator. The indices $A$ and $B$ are superindices and will be further explained and decomposed below. $J_{\sss AB}$ is graded antisymmetric while the superspace metric $I_{\sss AB}$ is graded symmetric and the induced scalar product between vectors is invariant under an $OSp(8^*|4)$ transformation, which can be seen as a definition of the supergroup.

When acting on fields, the generator $J_{\sss AB}$ is conventionally decomposed as
\beq
J_{\sss AB} = L_{\sss AB} + s_{\sss AB},
\eeq
where $L_{\sss AB}$ is the differential (orbital) piece and $s_{\sss AB}$ is the intrinsic (spin) piece. The latter acts only on the indices of the field.

By using the concept of triality in eight dimensions~\cite{Claus:1998,Arvidsson:2005}, we regard the superindex $A$ to be composed of $\alh=(1,\ldots,8)$, which is a chiral $SO(6,2)$ spinor index, and $a=(1,\ldots,4)$, which is a fundamental $USp(4)$ index (or equivalently, an $SO(5)$ spinor index). Furthermore, $\alh$ can naturally be decomposed into one chiral $SO(5,1)$ spinor index $\al=(1,\ldots,4)$ (subscript) and one anti-chiral $SO(5,1)$ spinor index $\al=(1,\ldots,4)$ (superscript). Using this decomposition, we may make contact with the usual notation for the superconformal group in six dimensions with $(2,0)$ supersymmetry. We find that the definitions
\beq
  J_{\sss AB} =
  \left( \begin{array}{ccc}
  \frac{1}{2} P_{\al\be} &  \frac{1}{2} \dia{M}{\al}{\be} + \frac{1}{4} \kde{\al}{\be} D & \frac{i}{2\sqrt{2}} Q^b_\al \\
  - \frac{1}{2} \dia{M}{\be}{\al} - \frac{1}{4} \kde{\be}{\al} D & - \frac{1}{2} K^{\al\be} & \frac{i}{2\sqrt{2}} \Om^{bc} S^\al_c \\
  - \frac{i}{2\sqrt{2}} Q^a_\be & - \frac{i}{2\sqrt{2}} \Om^{ac} S^\be_c & i U^{ab}
  \end{array} \right)
  \eqnlab{J_AB}
\eeq
and
\beq
  I_{\sss AB} =
  \left( \begin{array}{ccc}
  0 &  \kde{\al}{\be} & 0 \\
  \de^\al_{\ph{\al}\be} & 0 & 0 \\
  0 & 0 & i \Om^{ab}
  \end{array} \right),
  \eqnlab{superspacemetric}
\eeq
together with \Eqnref{superalgebra} reproduce all commutation relations of the six-dimensional superconformal group with conventions as in Ref.~\cite{Arvidsson:2005}. We have also introduced the $R$-symmetry invariant antisymmetric tensor $\Om^{ab}$ in the purely fermionic piece of the superspace metric.

It is convenient to define an inverse superspace metric, \ie a metric with superscript indices. This becomes
\beq
  I^{\sss AB} =
  \left( \begin{array}{ccc}
  0 &  \de^\al_{\ph{\al}\be} & 0 \\
  \kde{\al}{\be} & 0 & 0 \\
  0 & 0 & -i \Om_{ab}
  \end{array} \right),
  \eqnlab{superspacemetric_upper}
\eeq
which makes the relation
\beq
I_{\sss AB} I^{\sss BC} = \de^{\sss C}_{\sss A}
\eeq
valid (which is essential if we want to raise and lower indices). This requires that $\Om_{ab} \Om^{bc} = \kde{a}{c}$.

The next step is to introduce coordinates in the superconformal space. The most common choice would be to choose the vector representation of $SO(6,2)$ for the bosonic coordinates. However, we will instead use another eight-dimensional representation, the chiral spinor, for these coordinates. The fermionic coordinates are in the fundamental four-dimensional representation of $USp(4)$.

Thus, the coordinates in the superconformal space are denoted by $y_{\sss A}=(y_{\alh},y^a)=(y_\al,y^\al,y^a)$. It should be emphasized that $y_{\alh}$ are commuting (Grassmann even) while $y^a$ are anti-commuting (Grassmann odd). We also introduce a derivative $\pa_{\sss A}$ such that
\beq
\pa_{\sss A} y_{\sss B} = I_{\sss AB}.
\eeq
The use of a chiral spinor instead of the usual coordinate vector introduces a subtlety concerning reality. There is no Majorana-Weyl spinor in eight dimensions with signature $(6,2)$, which means that the components of $y_{\alh}$ cannot all be real~\cite{Kugo:1983}. The standard way of introducing a reality condition is instead to attach a fundamental $SU(2)$ index $i=(1,2)$ to $y_{\alh}$ and impose a symplectic $SU(2)$ Majorana condition. This additional $R$-symmetry can be motivated from the quaternionic structure of the conformal group $SO(6,2) \simeq SO(4;\mathbb{H})$. However, we will treat $y_{\alh}$ as an ordinary complex chiral spinor in this paper and not impose any reality condition.

In Ref.~\cite{Arvidsson:2005}, we related the coordinates $y_{\sss A}$ of the superconformal space to the conventional $(2,0)$ superspace coordinates $x^{\al\be}=-x^{\be\al}$ (six bosonic coordinates) and $\theta_a^\al$ (sixteen fermionic coordinates) by introducing a projective \emph{supercone}, inspired by Dirac's hypercone~\cite{Dirac:1936}. This supercone is defined by
\beq
\eqnlab{supercone}
y^2 \equiv I^{\sss AB} y_{\sss A} y_{\sss B} = 0,
\eeq
which clearly is invariant under $OSp(8^*|4)$ transformations. We will be interested in solutions to this condition of the form
\beq
\eqnlab{y-x}
\begin{aligned}
y^a &= \sqrt{2} \Om^{ab} \theta_b^\be y_\be \\
y^\al &= \left( 2 x^{\al\be} - i \Om^{ab} \theta^\al_a \theta^\be_b \right) y_\be,
\end{aligned}
\eeq
where the parameters $x^{\al\be}=-x^{\be\al}$ and $\theta^\al_a$ a priori need not be the corresponding coordinates in $(2,0)$ superspace. That this in fact is the case is indicated by considering their behavior under superconformal symmetry transformations; by requiring the left-hand and the right-hand sides of \Eqnref{y-x} to transform equally, we find that the parameters must transform according to the known transformation laws~\cite{Arvidsson:2005,Park:1998} for the corresponding $(2,0)$ superspace coordinates. Due to the projectiveness, we may always multiply the coordinate $y_{\sss A}$ by a constant and still remain on the supercone; this feature is obvious in \Eqnref{y-x} as well.

We should also mention that these relations look just like the super-twistor relations introduced by Ferber~\cite{Ferber:1978}. However, twistors~\cite{Penrose:1967,Penrose:1984,Penrose:1986} are conventionally used to describe the phase-space of on-shell particles while our aim is to consider space-time itself, without conjugate momenta. Twistors in six dimensions have been used in complex notation in Ref.~\cite{Hughston:1987} and in quaternionic notation in Ref.~\cite{Bengtsson:1988}. Our usage of twistors is rather similar to Witten's in the context of MHV amplitudes~\cite{Witten:2003,Sinkovics:2004}. We regard the coordinate $y_{\sss A}$ as a homogeneous coordinate in a projective super-twistor space. The latter space is a copy of the supermanifold $\mathbb{CP}^{7|4}$. We cannot take the coordinates to be real, but we can do the next best thing: we will only consider functions that depend on $y_{\alh}$ and $y^a$, not on their complex conjugates.

\section{Covariant fields}
\seclab{fields}

In this section, we will consider the $(2,0)$ tensor multiplet in six dimensions. Its bosonic content is a self-dual three-form gauge field $h_{\al\be}=h_{\be\al}$ and five scalar fields $\phi^{ab}$, transforming in the vector representation of $SO(5)$. This requires that $\Om_{ab} \phi^{ab} = 0$. The fermionic fields of the tensor multiplet are denoted $\psi^a_\al(x)$, transforming as chiral spinors under Lorentz transformations and in the ${\bf 4}$ representation of the $R$-symmetry group.

Since we are describing a supersymmetric field theory, it is convenient to work with superfields. For the $(2,0)$ tensor multiplet, there is an on-shell superfield formulation~\cite{Howe:1983} in terms of a superfield $\Phi^{ab}(x,\theta)$, defined over the $(2,0)$ superspace discussed in the previous section. Its use in our model is described thoroughly in Ref.~\cite{AFH:2003coupled}.

The superfield obeys the algebraic constraint $\Om_{ab} \Phi^{ab} = 0$, but also the differential constraint
\beq
\eqnlab{diff_con}
D^a_{\al} \Phi^{bc} + \frac{2}{5} \Om_{de} D^d_\al \left( \Om^{ab}
\Phi^{ec} + \Om^{ca} \Phi^{eb} + \frac{1}{2} \Om^{bc} \Phi^{ea}
\right) = 0,
\eeq
where $D^a_{\al}$ is the covariant superspace derivative, defined
according to
\beq
D^a_{\al}  = \pa^a_{\al} + i \Om^{ac} \theta_c^{\ga} \pa_{\al\ga}.
\eeq
In this expression and onwards, $\pa_{\al \be}$ denotes the derivative with respect to $x^{\al\be}$ while $\pa^a_\al$ is the derivative with respect to $\theta_a^\al$.

It is convenient to define supplementary superfields according to
\beq
\eqnlab{PsiH}
\begin{aligned}
\Psi^a_{\al} &= - \frac{2i}{5} \Om_{bc} D^b_\al \Phi^{ca} \\
H_{\al \be} &= \frac{1}{4} \Om_{ab} D^a_{\al} \Psi^b_{\be},
\end{aligned}
\eeq
but it should be noted that these contain no new degrees of freedom compared to $\Phi^{ab}$. The lowest components of the superfields  $\Phi^{ab}$, $\Psi^a_\al$ and $H_{\al\be}$ are the tensor multiplet fields $\phi^{ab}$, $\psi^a_\al$ and $h_{\al\be}$, hence the notation. The differential constraint \eqnref{diff_con} implies the free equations of motion for these fields.

After these preliminaries, we aim to construct a new superfield, living on the supercone defined in~\Eqnref{supercone}. This is supposed not only to transform linearly under superconformal transformations, but we also want it to contain all the fields of the tensor multiplet.

In order to incorporate self-duality in the eight-dimensional space in a simple way, we use Siegel's formalism~\cite{Siegel:1988,Siegel:1989,Siegel:1994}, where the manifestly covariant field corresponding to a three-form field strength in six dimensions is a four-form field. A self-dual four-form in eight dimensions transforms in the $\bf{35}_+$ representation of the Lorentz group. This representation can also be built from two symmetric chiral spinor indices, if we require tracelessness with respect to the eight-dimensional metric. Concretely, this means that we may write the self-dual four-form as $H_{\alh\beh}$. The tracelessness is accomplished by requiring that
\beq
I^{\alh\beh} H_{\alh\beh} = 0,
\eeq
where $I^{\alh\beh}$ is the purely bosonic piece of the metric $I^{\sss AB}$ in \Eqnref{superspacemetric_upper}.

Generalizing this to the superconformal space, it is natural to introduce a graded symmetric tensor field defined on the supercone, denoted by $\Ups_{\sss AB}(y)$. It should be supertraceless, meaning that
\beq
I^{\sss AB} \Ups_{\sss AB} = 0.
\eqnlab{supertraceless}
\eeq
In line with the general discussion by Mack and Salam~\cite{Mack:1969}, we demand the field to be a homogeneous function of $y$, \ie
\beq
\frac{1}{2} I^{\sss CD} y_{\sss C} \pa_{\sss D} \Ups_{\sss AB} = n \Ups_{\sss AB}.
\eqnlab{homo_Ups}
\eeq
We take the degree of homogeneity to be $n=-2$; this will be motivated later.

The superfield $\Ups_{\sss AB}$ also has to satisfy a subsidiary condition that reduces the number of degrees of freedom. The natural covariant expression that we impose on the field is
\beq
I^{\sss AB} y_{\sss A} \Ups_{\sss BC} =0,
\eqnlab{subcon}
\eeq
which should be valid on the entire supercone. If we expand this constraint by means of \Eqnref{y-x}, we find that
\beq
 \begin{aligned}
  \dia{\Ups}{\al}{\be} &= \left( 2 x^{\be\ga} + i \theta^\be \cdot \theta^\ga \right) \Ups_{\al\ga} - \sqrt{2}i\theta^\be_b \dia{\Ups}{\al}{b} \\
  \Ups^{a \be} &= \left( 2 x^{\be\ga} + i \theta^\be \cdot \theta^\ga \right) \dia{\Ups}{\ga}{a} + \sqrt{2} i \theta^\be_b \Ups^{ab} \\
  \Ups^{\al\be} &= \left( 2 x^{\be\ga} + i \theta^\be \cdot \theta^\ga \right) \dia{\Ups}{\ga}{\al} - \sqrt{2} i \theta_b^\be \Ups^{\al b},
 \end{aligned}
 \eqnlab{Ups-Ups}
\eeq
where the dot product between two $\theta$-coordinates is defined according to
\beq
\theta^\al \cdot \theta^\be \equiv \Om^{ab} \theta^\al_a \theta^\be_b,
\eeq
and this quantity is obviously symmetric in the spinor indices $\al$ and $\be$.

From \Eqnref{Ups-Ups}, we see that the most general solution to the algebraic equation \eqnref{subcon} is parametrized by the fields $\Ups_{\al\be}$, $\dia{\Ups}{\al}{b}$ and $\Ups^{ab}$. The supertracelessness condition \eqnref{supertraceless} becomes
\beq
\Om_{ab} \Ups^{ab} = 2 \theta^\al \cdot \theta^\be \Ups_{\al\be} - 2 \sqrt{2} \theta^\al_a \dia{\Ups}{\al}{a},
\eeq
and effectively removes one of the components in the parametrization fields. This suggests that we may use the superfields $\Phi^{ab}$, $\Psi^a_\al$ and $H_{\al\be}$ (or rather, the corresponding tensor multiplet fields, but it is more convenient to work with superfields) defined above to parametrize $\Ups_{\sss AB}$. However, from the supertracelessness condition it follows that we cannot simply identify these with the parametrization components in $\Ups_{\sss AB}$, since $\Phi^{ab}$ is supposed to be traceless.

The solution to this problem is to let
\beq
 \begin{aligned}
\Ups_{\al\be} &= \frac{1}{\ga^2} H_{\al\be} \\
\dia{\Ups}{\al}{b} &= \frac{1}{\ga^2} \left[ - \frac{3}{\sqrt{2}} \Psi^b_\al + \sqrt{2} \Om^{bc} \theta_c^\ga H_{\al\ga} \right] \\
\Ups^{ab} &= \frac{1}{\ga^2} \left[ -6i \Phi^{ab} - 3 \Om^{ac} \theta_c^\ga \Psi_\ga^b + 3 \Om^{bc} \theta_c^\ga \Psi_\ga^a + 2 \Om^{ac} \Om^{bd} \theta_c^\ga \theta_d^\de H_{\ga\de} \right],
 \end{aligned}
 \eqnlab{Ups_H1}
\eeq
where we have introduced factors of a projective parameter $\ga$ to take the degree of homogeneity into account~\cite{Mack:1969}.

The remaining components in $\Ups_{\sss AB}$ follow from \Eqnref{Ups-Ups} and are
\beq
 \begin{aligned}
\dia{\Ups}{\al}{\be} &= \frac{1}{\ga^2} \left[ (2x^{\be\ga} -i \theta^\be \cdot \theta^\ga ) H_{\al\ga} + 3i \theta^\be_c \Psi^c_\al \right] \\
\Ups^{\al\be} &= \frac{1}{\ga^2} \Big[  (2x^{\al\ga} -i \theta^\al \cdot \theta^\ga ) (2x^{\be\de} -i \theta^\be \cdot \theta^\de ) H_{\ga\de} - 12i \theta^\al_a \theta^\be_b \Phi^{ab} +  \\
& \quad + 6i \theta^{(\al}_a  (2x^{\be)\ga} -i \theta^{\be)} \cdot \theta^\ga ) \Psi^a_\ga \Big] \\
\Ups^{\al b} &= \frac{1}{\ga^2} \bigg[ (2x^{\al\ga} -i  \theta^\al \cdot \theta^\ga ) \left( \sqrt{2} \Om^{bd}\theta_d^\de H_{\ga\de} - \frac{3}{\sqrt{2}} \Psi^b_\ga \right) - {}  \\
& \quad - 6 \sqrt{2} \theta^\al_a \Phi^{ab} + 3 \sqrt{2} i \Om^{bc} \theta^\ga_c \theta^\al_a \Psi^a_\ga \bigg].
 \end{aligned}
 \eqnlab{Ups_H2}
\eeq
In this way, we have found a unique expression for the field $\Ups_{\sss AB}$ in terms of some fields $H_{\al\be}$, $\Psi^a_\al$ and $\Phi^{ab}$.

\Eqsref{Ups_H1} and \eqnref{Ups_H2} may be summarized in a very compact notation, where
\beq
\Ups_{\sss AB} = \frac{1}{\ga^2} \exp \left( -2 x^{\ga\de} s_{\ga\de} + 2 \sqrt{2} i \theta^\ga_d \dia{s}{\ga}{d} \right) \mathcal{H}_{\sss AB}.
\eeq
The matrix $\mathcal{H}_{\sss AB}$ is given by
\beq
\eqnlab{calH}
\mathcal{H}_{\sss AB} =
  \left( \begin{array}{ccc}
    H_{\al\be} & 0 & - \frac{3}{\sqrt{2}} \Psi^b_\al \\
    0 & 0 & 0 \\
    - \frac{3}{\sqrt{2}} \Psi^a_\be & 0 & -6i \Phi^{ab}
  \end{array} \right),
\eeq
and only contains the superfields in the $(2,0)$ tensor multiplet (no explicit coordinate dependence). Note that $\mathcal{H}_{\sss AB}$ is \emph{not} a covariant field in the superconformal space, it is just a compact way of collecting the superfields.

We also need the action of the intrinsic $OSp(8^*|4)$ generator $s_{\sss AB}$ on a field with two superindices; this is given by
\beq
\eqnlab{s_Ups}
s_{\sss CD} \Ups_{\sss AB} = \frac{1}{2} \left( I_{\sss DA} \Ups_{\sss CB} - (-1)^{\sss CD} I_{\sss CA} \Ups_{\sss DB} + (-1)^{\sss AB} I_{\sss DB} \Ups_{\sss CA} - (-1)^{\sss AB+CD} I_{\sss CB} \Ups_{\sss DA} \right)
\eeq
and is consistent with the commutation relations \eqnref{superalgebra}.

Since the commutators $[x\cdot s,x\cdot s]$, $[x\cdot s, \theta \cdot s]$ and $[\theta \cdot s,\theta \cdot s]$ all are zero, we may also write the inverse relation
\beq
\mathcal{H}_{\sss AB} = \ga^2 \exp \left( 2 x^{\ga\de} s_{\ga\de} - 2 \sqrt{2} i \theta^\ga_d \dia{s}{\ga}{d} \right) \Ups_{\sss AB}.
\eqnlab{H_Ups}
\eeq
This is an important result, stating how to extract the $(2,0)$ tensor multiplet fields from a manifestly superconformally covariant field on the supercone. Alternatively, it shows how to build a covariant field from the tensor multiplet superfields. If we compare this to the general recipe by Mack and Salam~\cite{Mack:1969}, we note that the relations between the fields are very similar. In that paper, the fields in the conformal space are multiplied by an operator $V(x)$, with the purpose of making the generator of translations act only differentially (without any intrinsic piece) on the fields in six dimensions. In our case, this operator is replaced by
\beq
V(x,\theta)= \exp\left(2 x^{\ga\de} s_{\ga\de} - 2 \sqrt{2} i \theta^\ga_d \dia{s}{\ga}{d} \right),
\eeq
which makes both translations and supersymmetry transformations act only differentially on the $(2,0)$ superfields.

Following Mack and Salam~\cite{Mack:1969}, this suggests a way of transforming the other generators of superconformal transformations according to
\beq
J_{\sss AB} \rightarrow V(x,\theta) J_{\sss AB} V^{-1}(x,\theta).
\eeq
This gives the explicit generators acting on $\mathcal{H}_{\sss AB}$, and the corresponding transformations laws may be found rather easily. These are found to agree with the known~\cite{Arvidsson:2005} transformation laws for the superfields forming the $(2,0)$ tensor multiplet, showing that our anticipation was correct --- the fields used to parametrize the solution to the algebraic constraint \eqnref{subcon} may consistently be interpreted as the superfields of the $(2,0)$ tensor multiplet. This derivation also yields a nice insight into the origins of the different pieces of the transformation laws.

\section{The differential constraint}
\seclab{diff_con}

What is the purpose of finding a manifestly covariant formalism? An obvious advantage of such a formulation is the possibility to find covariant quantities and transformation laws in a simple way. For example, if we write down a scalar in the superconformal space composed of covariant quantities, we know that its corresponding field in the ordinary superspace will be invariant (in a certain sense, see below) under superconformal transformations.

The simplest non-zero $OSp(8^*|4)$ scalar that we may form from our ingredients is quadratic in the field $\Ups_{\sss AB}(y)$ and written as
\beq
I^{\sss AD} I^{\sss BC} \Ups_{\sss AB} \Ups_{\sss CD} = -36 \frac{1}{\ga^4} \Om_{ac} \Om_{bd} \Phi^{ab} \Phi^{cd},
\eeq
where we used \Eqsref{Ups_H1} and \eqnref{Ups_H2} to translate the fields $\Ups_{\sss AB}(y)$ to $(2,0)$ superfields. We recognize the right-hand side of this equation as the scalar related to the string tension when the tensor multiplet is coupled to a self-dual string~\cite{AFH:2003coupled}. This shows that this quantity, designed to be invariant under supersymmetry, is a superconformal scalar.

Usually, a scalar field transforms only differentially, but in the case of superconformal scalars we have to include the homogeneity degree as well. This means that
\beq
\de \left( \Om_{ac} \Om_{bd} \Phi^{ab} \Phi^{cd} \right) = \left[\de x \cdot \pa + \de \theta \cdot \pa + 4 \La(x,\theta) \right] \left( \Om_{ac} \Om_{bd} \Phi^{ab} \Phi^{cd} \right),
\eeq
in agreement with the transformation laws in Ref.~\cite{Arvidsson:2005}. $\La(x,\theta)\equiv\la-2 c \cdot x + i\rho \cdot \theta$ is a superspace-dependent parameter function, including the parameters for dilatations, special conformal transformations and special supersymmetry transformations.

Let us move on to the main purpose of this section: to investigate whether the differential constraint~\eqnref{diff_con} for the superfield $\Phi^{ab}(x,\theta)$ may be formulated in a manifestly covariant way, with respect to superconformal symmetry. We expect this to be possible, since the differential constraint respects superconformal symmetry and is formulated in terms of superfields and superderivatives.

We do not have very many quantities to build such a covariant constraint from. Considering what the differential constraint and the derived constraints in Ref.~\cite{AFH:2003coupled} look like, we expect the covariant expression to have four free superindices.

Firstly, note that the graded symmetry of $I_{\sss AB}$ and \Eqnref{s_Ups} together imply that
\beq
\eqnlab{s_Ups=0}
s_{\sss [AB} \Ups_{\sss C]D} = 0,
\eeq
where, of course, the antisymmetrization is graded.

Having done this observation, it makes sense to consider the related equation
\beq
L_{\sss [AB} \Ups_{\sss C]D} = 0,
\eeq
which is a differential analogue of \Eqnref{s_Ups=0}. Using the expressions \eqnref{Ups_H1} and \eqnref{Ups_H2} for $\Ups_{\sss AB}(y)$ together with explicit expressions for the different pieces of $L_{\sss AB}$, we find that the equation is satisfied exactly when the superfield $\Phi^{ab}(x,\theta)$ obeys the differential constraint~\eqnref{diff_con}, but only if the degree of homogeneity of $\Ups_{\sss AB}$ is $n=-2$. The latter observation is interesting and shows that $n$ cannot be chosen freely.

This means that we may indeed formulate a manifestly superconformally covariant differential constraint, namely
\beq
J_{\sss [AB} \Ups_{\sss C]D} = 0.
\eqnlab{diff_con_super}
\eeq
This also implies that the tensor multiplet fields must obey the free equations of motion, since they are a consequence of the differential constraint. So, the constraint~\eqnref{diff_con_super} is both a constraint on the superfield $\Ups_{\sss AB}$ and an equation of motion.

This important result is quite remarkable --- \Eqnref{diff_con_super} contains a lot of information about the tensor multiplet fields in a very simple and manifestly superconformally covariant formulation.

To conclude: we have constructed a manifestly superconformally covariant formulation of the six-dimensional $(2,0)$ tensor multiplet, by considering a graded symmetric superfield on a projective supercone in a superspace with eight bosonic and four fermionic dimensions. We have also found a covariant expression for the differential constraint, which is essential for the consistency of the conventional $(2,0)$ superfield formulation. Hopefully, this formalism may be applied to other problems in this theory, as a useful tool to include superconformal covariance in a manifest way. It should also be possible to generalize these concepts to other superconformal theories.

\vspace{7mm}
\noindent
\textbf{Acknowledgments:}
I would like to thank Erik Flink and M{\aa}ns Henningson for
stimulating and encouraging discussions. I am very grateful to Robert Marnelius for introducing me to Dirac's formalism. I would also like to thank Martin Cederwall and Bengt E.W.~Nilsson for discussions on twistors.

%\clearpage
\bibliographystyle{utphysmod3b}
\addcontentsline{toc}{section}{References}
\bibliography{biblio}

\end{document}

%% file: macros.tex
\newcommand {\beq} {\begin{equation}}
\newcommand {\eeq} {\end{equation}}
\newcommand {\beqa} {\begin{eqnarray}}
\newcommand {\eeqa} {\end{eqnarray}}
\newcommand {\beqan} {\begin{eqnarray*}}
\newcommand {\eeqan} {\end{eqnarray*}}

\newcommand {\ph}[1]{\phantom{#1}}
\newcommand {\ie}{i.e.~}

\newcommand {\sss} {\scriptscriptstyle}

\newcommand{\al}{\ensuremath{\alpha}}
\newcommand{\be}{\ensuremath{\beta}}
\newcommand{\ga}{\ensuremath{\gamma}}

\newcommand{\de}{\ensuremath{\delta}}

\newcommand{\la}{\ensuremath{\lambda}}

\newcommand{\La}{\ensuremath{\Lambda}}

\newcommand{\Om}{\ensuremath{\Omega}}

\newcommand{\Ups}{\ensuremath{\Upsilon}}

\newcommand{\alh}{\ensuremath{\hat{\alpha}}}
\newcommand{\beh}{\ensuremath{\hat{\beta}}}

\newcommand{\kde}[2]{\de_{#1}^{\ph{#1}#2}}
\newcommand{\dia}[3]{{#1}_{#2}^{\ph{#2}#3}}

\newcommand{\mathHb}[1]{{\mathop{\kern0pt#1}\limits^{\,\sss
      \prime\prime}\vphantom{#1}}}

\newcommand {\pa} {\partial}

\newcommand{\eqnlab}[1]{\label{eqn:#1}}

\newcommand{\eqnref}[1]{(\ref{eqn:#1})}

\newcommand{\Eqnref}[1]{Eq.~(\ref{eqn:#1})}

\newcommand{\Eqsref}[1]{Eqs.~(\ref{eqn:#1})}

\newcommand{\seclab}[1]{\label{sec:#1}}

\newcommand{\Secref}[1]{Section~\ref{sec:#1}}

%% file: supermanifest_v5.bbl
\providecommand{\href}[2]{#2}\begingroup\raggedright\begin{thebibliography}{10}

\bibitem{Dirac:1936}
P.~A.~M. Dirac,  {\em Wave equations in conformal space}, Ann. Math. {\bf 37}
  (1936) 429.

\bibitem{Kastrup:1966}
H.~A. Kastrup,  {\em Gauge properties of the {M}inkowski space}, Phys. Rev.
  {\bf 150} (1966) 1183.

\bibitem{Marnelius:1979}
R.~Marnelius and B.~E.~W. Nilsson,  {\em Equivalence between a massive spinning
  particle in {M}inkowski space and a massless one in a de {S}itter space},
  Phys. Rev. {\bf D20} (1979)
839.
%%CITATION = PHRVA,D20,839;%%.

\bibitem{Marnelius:1979b}
R.~Marnelius,  {\em Manifestly conformally covariant description of spinning
  and charged particles}, Phys. Rev. {\bf D20} (1979)
2091.
%%CITATION = PHRVA,D20,2091;%%.

\bibitem{Marnelius:1980}
R.~Marnelius and B.~E.~W. Nilsson,  {\em Manifestly conformally covariant field
  equations and a geometrical understanding of mass}, Phys. Rev. {\bf D22}
  (1980)
830.
%%CITATION = PHRVA,D22,830;%%.

\bibitem{Mack:1969}
G.~Mack and A.~Salam,  {\em Finite component field representations of the
  conformal group}, Ann. Phys. {\bf 53} (1969)
174--202.
%%CITATION = APNYA,53,174;%%.

\bibitem{Nahm:1978}
W.~Nahm,  {\em Supersymmetries and their representations}, Nucl. Phys. {\bf
  B135} (1978)
149.
%%CITATION = NUPHA,B135,149;%%.

\bibitem{Witten:1995}
E.~Witten,  {\em Some comments on string dynamics},
\href{http://arXiv.org/abs/hep-th/9507121}{{\tt hep-th/9507121}}.
%%CITATION = HEP-TH 9507121;%%.

\bibitem{Arvidsson:2005}
P.~Arvidsson,  {\em Superconformal symmetry in the interacting theory of (2,0)
  tensor multiplets and self-dual strings}, J. Math. Phys. {\bf 47} (2006)
  042301
[\href{http://www.arXiv.org/abs/hep-th/0505197}{{\tt hep-th/0505197}}].
%%CITATION = HEP-TH 0505197;%%.

\bibitem{Kac:1977}
V.~G. Kac,  {\em A sketch of {L}ie superalgebra theory}, Commun. Math. Phys.
  {\bf 53} (1977)
31--64.
%%CITATION = CMPHA,53,31;%%.

\bibitem{Claus:1998}
P.~Claus, R.~Kallosh and A.~Van~Proeyen,  {\em M5-brane and superconformal
  (0,2) tensor multiplet in 6 dimensions}, Nucl. Phys. {\bf B518} (1998)
  117--150
[\href{http://www.arXiv.org/abs/hep-th/9711161}{{\tt hep-th/9711161}}].
%%CITATION = HEP-TH 9711161;%%.

\bibitem{Kugo:1983}
T.~Kugo and P.~K. Townsend,  {\em Supersymmetry and the division algebras},
  Nucl. Phys. {\bf B221} (1983)
357.
%%CITATION = NUPHA,B221,357;%%.

\bibitem{Park:1998}
J.-H. Park,  {\em Superconformal symmetry in six-dimensions and its reduction
  to four}, Nucl. Phys. {\bf B539} (1999) 599--642
[\href{http://www.arXiv.org/abs/hep-th/9807186}{{\tt hep-th/9807186}}].
%%CITATION = HEP-TH 9807186;%%.

\bibitem{Ferber:1978}
A.~Ferber,  {\em Supertwistors and conformal supersymmetry}, Nucl. Phys. {\bf
  B132} (1978)
55.
%%CITATION = NUPHA,B132,55;%%.

\bibitem{Penrose:1967}
R.~Penrose,  {\em Twistor algebra}, J. Math. Phys. {\bf 8} (1967)
345.
%%CITATION = JMAPA,8,345;%%.

\bibitem{Penrose:1984}
R.~Penrose and W.~Rindler, {\em Spinors and space-time. Vol. 1: Two-spinor
  calculus and relativistic fields}.
\newblock Cambridge University Press, 1984.

\bibitem{Penrose:1986}
R.~Penrose and W.~Rindler, {\em Spinors and space-time. Vol. 2: Spinor and
  twistor methods in space-time geometry}.
\newblock Cambridge University Press, 1986.

\bibitem{Hughston:1987}
L.~P. Hughston and W.~T. Shaw,  {\em Minimal curves in six dimensions}, Class.
  Quant. Grav. {\bf 4} (1987)
869.
%%CITATION = CQGRD,4,869;%%.

\bibitem{Bengtsson:1988}
I.~Bengtsson and M.~Cederwall,  {\em Particles, twistors and the division
  algebras}, Nucl. Phys. {\bf B302} (1988)
81.
%%CITATION = NUPHA,B302,81;%%.

\bibitem{Witten:2003}
E.~Witten,  {\em Perturbative gauge theory as a string theory in twistor
  space}, Commun. Math. Phys. {\bf 252} (2004) 189--258
[\href{http://www.arXiv.org/abs/hep-th/0312171}{{\tt hep-th/0312171}}].
%%CITATION = HEP-TH 0312171;%%.

\bibitem{Sinkovics:2004}
A.~Sinkovics and E.~Verlinde,  {\em A six dimensional view on twistors}, Phys.
  Lett. {\bf B608} (2005) 142--150
[\href{http://www.arXiv.org/abs/hep-th/0410014}{{\tt hep-th/0410014}}].
%%CITATION = HEP-TH 0410014;%%.

\bibitem{Howe:1983}
P.~S. Howe, G.~Sierra and P.~K. Townsend,  {\em Supersymmetry in six
  dimensions}, Nucl. Phys. {\bf B221} (1983)
331.
%%CITATION = NUPHA,B221,331;%%.

\bibitem{AFH:2003coupled}
P.~Arvidsson, E.~Flink and M.~Henningson,  {\em Supersymmetric coupling of a
  self-dual string to a (2,0) tensor multiplet background}, J. High Energy
  Phys. {\bf 11} (2003) 015
[\href{http://www.arXiv.org/abs/hep-th/0309244}{{\tt hep-th/0309244}}].
%%CITATION = HEP-TH 0309244;%%.

\bibitem{Siegel:1988}
W.~Siegel,  {\em Conformal invariance of extended spinning particle mechanics},
  Int. J. Mod. Phys. {\bf A3} (1988)
2713--2718.
%%CITATION = IMPAE,A3,2713;%%.

\bibitem{Siegel:1989}
W.~Siegel,  {\em All free conformal representations in all dimensions}, Int. J.
  Mod. Phys. {\bf A4} (1989)
2015.
%%CITATION = IMPAE,A4,2015;%%.

\bibitem{Siegel:1994}
W.~Siegel,  {\em Supermulti - instantons in conformal chiral superspace}, Phys.
  Rev. {\bf D52} (1995) 1042--1050
[\href{http://www.arXiv.org/abs/hep-th/9412011}{{\tt hep-th/9412011}}].
%%CITATION = HEP-TH 9412011;%%.

\end{thebibliography}\endgroup
